# Sarbanes-Oxley: What About all the Spreadsheets?

## Controlling for Errors and Fraud in Financial Reporting

**Raymond R. Panko & Nicholas Ordway**

Raymond R. Panko
University of Hawai'i
panko@hawaii.edu

Nicholas Ordway
University of Hawai'i
olgierd@hawaii.edu

### Abstract

The Sarbanes-Oxley Act of 2002 has finally forced corporations to examine the validity of their spreadsheets. They are beginning to understand the spreadsheet error literature, including what it tells them about the need for comprehensive spreadsheet testing. However, controlling for fraud will require a completely new set of capabilities, and a great deal of new research will be needed to develop fraud control capabilities. This paper discusses the riskiness of spreadsheets, which can now be quantified to a considerable degree. It then discusses how to use control frameworks to reduce the dangers created by spreadsheets. It focuses especially on testing, which appears to be the most crucial element in spreadsheet controls.

### Keywords

Cell error rate (CER). CobiT. controls. control deficiency. control framework. COSO. end-user computing (EUC). error. error rate floor. fraud. 17799. ITIL. material. spreadsheet. testing.

### Figures

Figures are printed at the end of the paper.

# Introduction

## Sarbanes-Oxley

In 2002, after financial reporting frauds at Enron and other major companies, the U.S. Congress passed the Sarbanes-Oxley Act. This act covers many things, including the retention of documents. However, the focus of the Act, which is often called SOX, is Section 404. This section requires chief corporate officers to assess whether their financial reporting systems are effective. Furthermore, it assumes an independent external auditor must assess the officers' assessment.

To implement SOX, Congress created the Public Company Accounting Oversight Board (PCAOB) to create auditing standards. PCAOB's main guidance on 404 assessments of control attestations has been Auditing Standard No. 2, An *Audit of Internal Control Over Financial Reporting Performed in Conjunction with an Audit of Financial Statements.*

The focus of SOX and of Auditing Standard 2 is the creation of effective controls. Figure 1 illustrates that controls are ways to help a corporation achieve its goals, such as producing accurate financial reports. Controls cannot guarantee that the goals will be met, but they reduce the risk that these goals will not be met. In this context, effectively controlled financial reporting processes give *reasonable* assurance that the company will meet the goal of producing accurate financial reports.

**Effectively controlled financial reporting processes give *reasonable* assurance that the company will meet the goal of producing accurate financial reports.**

**Figure 1: Controls**

According to Standard 2, an internal control deficiency exists when the design or operation of a control does not allow for the timely prevention or detection of misstatements. The standard defines two types of deficiencies:

- In a *significant* deficiency, there is more than a remote likelihood that the financial statements will be impacted in a manner that is consequential but not material.

- In a *material* deficiency, there is "a significant deficiency, or combination of significant deficiencies, that results in more than a remote likelihood that a material misstatement of the annual or interim financial statements will not be prevented or detected." Vorhies [2005] indicates that a 5% error in revenues is the usual threshold for labeling a deficiency as material because a smaller difference is not likely to sway a reasonable investor.

This distinction between significant and material internal control deficiencies is important because if management finds even a single material deficiency, it is not allowed to assess its internal controls as effective.

## The Use of Spreadsheets in Financial Reporting

Auditing Standard No. 2 clarifies that controls must involve all forms of information technology (IT) used in financial reporting. One particular IT concern for corporations is the use of spreadsheets in financial reporting. There have long been indications that many spreadsheets are large [Cale, 1994; Cragg & King, 1993; Floyd, Walls, & Marr, 1995; Hall, 1996], complex [Hall, 1996], and very important to their firm [Chan & Storey, 1996; Gable, Yap & Eng, 1991; Hall, 1996;]. When Comshare, Inc. surveyed 700 finance and budgeting professionals in the mid- I 990s, it found that spreadsheets were already dominating budgeting [Modem Office Technology, 1994]. In turn, when CPS [2001] did a survey of strategic planning practices in major U.S. corporation, 72% of respondents relied on spreadsheets exclusively to do strategic planning.

Although some people may doubt that companies use spreadsheets in critical financial reporting operations, the widespread use of spreadsheets is well documented.

- Financial intelligence firm CODA reports that 95% of U.S. firms use spreadsheets for financial reporting according to its experience (www.coda.com).

- In 2004, RevenueRecognition.com [2004] (now Softtrax) had the International Data Corporation interview 118 business leaders. IDC found that 8 5% were using spreadsheets in financial reporting and forecasting.

- CFO.com [Durfee 2004] interviewed 168 finance executives in 2004. The interviews asked about infori-nation technology use in the finance department. Out of *14* technologies discussed, only two were widely used-spreadsheets and basic budgeting and planning systems. Every subject said that his or her department used spreadsheets.

- In 2003, the Hacket Group (www.thehacketgroup.com) surveyed mid-size companies. It found that 47% of companies use stand-alone spreadsheets for planning and budgeting.

- In Europe, A.R.C. Morgan interviewed 376 individuals responsible for overseeing SOX compliance in multinationals that do business in the United States [TMCnet.com, *2004].* These respondents came from *21* different countries. More than 80% of the respondents said that their firms use spreadsheets for both managing the control environment and financial reporting.

- In a webcast for Deloitte on May *22, 2005,* the first author was able to ask a series of questions of the audience. The average response size was just over 800 financial professionals and officers in corporations. One question specifically asked, "Does your firm use spreadsheets of *material importance* in financial reporting?" Of the respondents, 87.7% answered in affirmative., while 7.1% said, "No." (Another *5.2%* chose "Not Applicable.")

Furthermore, when companies use spreadsheets for financial reporting, they often use many. One firm used more than 200 spreadsheets in its financial planning process.

Today, companies are widely confused over what to do about spreadsheet errors. Obviously, if financial reporting spreadsheets contain a significant number of errors and a reasonable amount of testing has not been done, it is difficult to say that the reporting process is well controlled.

Most firms will have to implement Section 404 for their Year-End 2005 financial reports. Most firms and external auditing companies are likely to issue vague statements about their spreadsheet controls or are likely to ignore spreadsheets entirely. However, as we will see below, these assessments are likely to be wrong.

## Lack of Controls

One concern with spreadsheets is that they are rarely well controlled [Davies & Ikin, 1987; Cragg & King, 1993; Fernandez, 2002; Floyd, Walls, & Marr, 1995; Gosling, 2003; Hall, 1996; Hendry & Green, 1994; Nardi & Miller, 1991; Nardi, 1993; Schultheis & Sumner, 1994). This is not surprising because few organizations have serious control policies-or indeed any policies at all-for spreadsheet development [Cale, 1994; Fernandez, 2002; Floyd, Walls, & Marr, 1995; Galletta & Hufnagel, 1992; Hall, 1996; Speier & Brown, 1996].

A specific concern is testing. Although there has long been evidence that spreadsheet error, at least, is widespread, organizations rarely mandate that spreadsheets and other end user applications be tested after development [Cale, 1994; Cragg & King, 1993; Floyd, Walls, & Marr, 1995; Galletta & Hufnagel, 1992; Gosling, 2003; Hall, 1996; Speier & Brown, 1996]. Also, individual developers rarely engage in systematic testing on their own spreadsheets after development [Cragg & King, 1993; Davies & Ikin, 1987; Hall, 1996; Schultheis & Sumner, 1994].

As noted earlier, the first author was able to ask questions of corporate financial professionals and officers in a webcast. Figure 2 shows respondent answers when they were asked, "For spreadsheets of material importance used in financial reporting, what percentage does your company test?" By and large, this figure shows that few spreadsheets are tested, even if they are of material importance. However, it shows that 17% of the respondents said that their firm tests more than 25% of their material financial spreadsheets, and 16% said that their firm tests nearly all. This makes it appear that a sizeable minority of companies do test their spreadsheets. However, as we will see later, what most respondents call testing is "looking over the spreadsheet," not comprehensive cell-by-cell testing.

**Figure 2: Testing for Material Financial Spreadsheets**

This lack of comprehensive testing may exist because developers tend to be overconfident of the accuracy of their untested spreadsheets. Certainly, widespread overconfidence, often in the face of widespread errors, has been seen repeatedly in spreadsheet research [Brown & Gould, 1987; Davies & lkin, 1987; Floyd, Walls, & Marr, 1995; Panko & Halverson, 1997; Panko & Featherman, 2005].

In a vicious cycle, organizations that do not test their spreadsheets get no feedback on real error rates and so do not realize the ubiquity of spreadsheet errors, and, therefore, see no need for testing. Rasmussen [ 1974] has noted that people use stopping rules to decide when to stop doing activities such as testing. If people are overconfident they are likely to stop too early. Consequently, if firms use spreadsheets to make decisions but do not test their spreadsheets, they may not realize how many errors there are in their spreadsheets.

One might argue that the real world would provide painful feedback if a spreadsheet were incorrect. For some situations, such as budgeting, errors would have to be small in order to pass undetected. Unfortunately, in this case, even small percentage errors can be very damaging. Hicks [ 1995] found that a relatively small percentage error in the capital budgeting spreadsheet he examined would have produced an error of several hundred million dollars. At the other extreme, when a new situation is modeled, such as the purchase of another company, even large errors in the spreadsheet might not be obvious. If a promising corporate purchase goes bad, furthermore, it would be easy to dismiss the problem as being due to unforeseen factors even if the real problem is a spreadsheet error. Without testing, real-world feedback may not be very effective.

Overall, spreadsheet errors are widespread and there are strong prospects for fraud This situation is at odds with the looseness with which most organizations treat their spreadsheets.

## The Prevalence of Spreadsheet Errors

One problem with spreadsheets is the specter of errors. If even a single spreadsheet has more than a remote likelihood of creating a material financial error through having errors, the organization will not be able to assess its entire financial reporting system as effective. In addition, if a company uses a large number of spreadsheets in financial reporting-as most firms do-then a collection of spreadsheets that individually only have significant potential deficiencies because of errors may constitute a material internal control deficiency.

Are errors common in spreadsheets? The answer is that errors are extremely common. The most convincing data on this come from audits of real-world operational spreadsheets. Figure 3, which presents data from several audit studies, shows convincingly that spreadsheet errors are anything but rare.

**Figure 3: Audits of Real-World Spreadsheets**

- First, these audits found errors in nearly all (94%) of the spreadsheets they audited. This percentage would have been even higher, but several of the studies only reported serious errors. In other words, we can expect nearly all spreadsheets to contain errors.

- Second, these audits found many errors in the spreadsheets they audited.

- Specifically, studies that measured errors on a per-formula basis [Butler, 2000; Clermont, Hanin, & Mittermeier, 2002; Hicks. 1995- Lawrence & Lee, 2004; Lukasic, 1998] found errors in 5.2% of the formulas in these spreadsheets. Most large spreadsheets have thousands of formula cells, so most large spreadsheets probably have dozens or even hundreds of errors.

If this cell error rate seems excessive, it should not. There has been a great deal of research on human error [Panko, 2005a], and for tasks of comparable complexity, such as writing computer program statements, similar error rates are also seen. Panko (2005a) has summarized results from a number of studies that measured fault rates in real-world software. Of particular value are four large studies [Madachy, 1996; O'Neill, 1994; Ebenau & Strauss, 1994; Weller 19931. In these studies, average fault rates ranged from 1.5% to 2.6%. Note that this is very close to the cell error rates seen in Figure 3 for spreadsheet code inspection. Zage and Zage [1993] and Grady [1992] both found that fault rates depend on program difficulty. In both studies, fault rates were at least twice as high for difficult programs as for simple programs.

Humans appear to have an error rate floor that exists despite working very carefully. Everyone has a similar error rate floor, and working more carefully can decrease this error rate only modestly. Research has shown that the same human cognitive processes that allow us to respond to the world correctly most of the time have unavoidable trade-offs that create errors a few percent of the time [Reason, 19901. In most human cognitive activities, such small error rates are only minor nuisances, if anyone notices them at all. However, when dozens of formula cells are on a chain to calculate a bottom-line financial value, the probability of error in the bottom-line value becomes unacceptable,

How many spreadsheet errors are significant or material? (As noted earlier, a 5% error in an important bottom-line value in a key financial variable would probably be considered a material error [Vorhies, 2005].) Unfortunately, we lack hard data on this critical issue. However, when Panko [2005b] interviewed the two spreadsheet auditing principals, both independently gave data suggesting that about 5% of all spreadsheets contain what one of the interviewees called "show stopper" errors. In addition, the Coopers and Lybrand [ 1997] study shown in Figure 3 did not report an error unless there was at least a 5 % error in a bottom line value. The study found such errors in 9 1 % of all spreadsheets. KPMG found a similar error rate and only reported spreadsheets to be incorrect if they contained errors that would make a difference to decision makers.

We also have data from software inspections, which sometimes classify errors as major or minor. Although definitions about what constitutes a major error differ, all software audit studies that have used the major/minor distinction found that major errors are very common. Bush [ 1994] and Jones [ 1996] both reported that a quarter of the errors in the inspections they examined were major errors. O'Neill [1994] found only 13% of errors to be major, while Schulmeyer [ 1999] found that 42% of all errors were ma or. In addition, Ebenau and Strauss [ 1994] and Weller [ 1993 ] found maj or errors in 1.4% to 2.8% of the lines of code examined but did not report major errors as a percentage of total errors. Given this data from software inspections, it certainly would be risky to assume that nearly all spreadsheet errors will be minor.

## The Prospect of Spreadsheet Fraud

Although there has been a great deal of research on spreadsheet error, there has not been formal research on spreadsheet fraud. Legally, a fraud exists when one person knowingly lies or conceals information whose nondisclosure would make other statements misleading, in order to get the victim to act in a way contrary to the victim's interests, Note that two elements are needed for there to be fraud: deception and harm,

There has long been concern that spreadsheet developers will use optimistic (or pessimistic) assumptions to make their plans and actions look better [Levy, 1984]. However, few people would consider this "puffery" to be fraud because people who read spreadsheets should be somewhat skeptical consumers of spreadsheet results. However, if people to whom spreadsheets results are submitted are not likely to be aware of the degree of risk, this is at least an ethical violation [Newell, 1995]. In addition, when the damage due to deception increases in intensity, then spreadsheet misanalysis rises to the level of fraud.

## H. M. Customs and Excise

Spreadsheet fraud is not just a theoretical concern. In England, the H. M. Customs and Excise collects certain types of taxes. When spreadsheets became prevalent in tax submissions, the agency began to audit submitted spreadsheets and found that many had substantial errors. In the late 1990s, the agency developed a program to automate many aspects of this auditing process [Butler, 2000]. This was SpACE (Spreadsheet Auditing by Customs & Excise). In addition to looking for innocent errors, the program also looked for certain types of fraud that the agency had found in earlier audits. For instance, the program looks for a number that was entered as text. Excel treats text cells as having the value zero. Consequently, entering a number in a column of numbers as text reduces the real total. This will create a fraudulent reduction in tax payments.

## The All First Fraud Scandal

The most famous case of spreadsheet fraud occurred at Allfirst, which was a U.S. subsidiary of Allied Irish Banks (AIB) in Ireland. The best information about this case comes from the United States Department of Justice (2002).

A currency trader, John Rusnak, began losing money in his trades around 1997. He used a series of spreadsheet subterfuges to hide his losses, which continued to increase, When the fraud was finally discovered, his losses amounted to $691.2 million USD. Although neither Allfirst nor Allied Irish Bank went into receivership, the losses amounted to 60% of AIB's 2001 revenues, and losses produced major drops in stock prices. After the scandal, AIB sold off its Allfirst subsidiary.

AIB immediately commissioned Eugene Ludwig, former U.S. Comptroller of the Currency, to prepare a report on the incident. The story his report tells is an excellent cautionary tale.

Rusnak began by entering two false option trades in the company's trading system-receipt of a large premium and the payment of a large premium. The first option would expire the day of the trade, the other one later. Allfirst had no reports on options that expired the same day and so did not detect what Rusnak was doing. The second option created a false asset on the company's books. This offset the real losing position that Rusnak wished to hide.

Initially, Rusnak used fake broker confirmations to validate his fictitious deals. This was risky because the back office staff reconciled trades with receipts. However, Rusnak convinced back-office personnel that they did not have to confirm the trades because they were offsetting deals with no transfer of cash.

In 2001, the head of treasury funds at Allfirst noted that Rusnak's trades were using up an unusually large portion of his balance sheet and that this was disproportionate to his earnings. He ordered Rusnak to reduce his exposure on the balance sheet. Rusnak did so, but he accomplished this using highly risk trades that saddled the company with massive potential liabilities.

One control at Allfirst was to compute a value-at-risk (VaR) ratio for each trader. The data for these calculations were supposed to have been computed independent by the back office staff, but Rusnak was able to persuade them to use data on his computer. He manipulated this data to make his VaR ratio look acceptable.

The fraud came apart when a back office supervisor noticed that Rusnak's trades were not being confirmed as required by procedures. The supervisor discovered that a number of trades were clearly bogus. He notified management of the problem. The fraud quickly unraveled.

Rusnak eventually entered into a plea agreement that got him only seven years in jail. This relatively "light" sentence was a result of his agreeing to work with prosecutors in prosecuting people in other companies whose actions prolonged the time it took for Rusnak's scheme to unravel [BBC, 2002].

# Legislation: SOX and Beyond

## Sarbanes-Oxley (SOX) and Financial Fraud

We have seen that the Sarbanes-Oxley Act of 2002, also known as SOX, requires U.S. firms and the many foreign firms listed on U.S. stock exchanges to have effective controls for their financial reporting systems and to report on the effectiveness of these controls. For most large firms, the deadline to implement effective controls either has passed or will pass soon.

SOX gives the Securities and Exchange Commission (SEC) overall responsibility for implementing the law. The SEC, in turn, created the Public Company Accounting Oversight Board (PCAOB) to develop specific rules and oversight functions to implement Sarbanes-Oxley.

Although SOX was a response to specific high-profile cases of fraud, fraud in financial reporting has long been a major problem. In 2004, fraud through financial statements represented only 7% of all fraud cases studied in an ACFE survey [ACFE, 2004], but the median fraud loss for financial fraud was a million dollars, compared to a median loss of only $100,000 for frauds in general.

## Other Regulations

Although SOX has received the most attention, a number of other recent pieces of legislation have required corporations to reconsider their financial systems and other information systems.

**SEC Accelerated Filing Deadlines**

Beginning in December 2002, the SEC has required firms to reduce the time they may take to produce their quarterly and annual reports. These tighter time limits will make designing controls more, difficult because there will be less time to check for errors and violations.

**IAS/IFRS**

In accounting, U.S. accounting standards are set by the U.S. Federal Accounting Standards Board (FASB), which creates the generally accepted accounting practices (GAAP). In Europe, the International Accounting Standards (lAS) will have to be followed by U.S. firms operating in Europe, including the finance-specific International Financial Reporting Standards (IFRS).

**SAS 99**

In 2002, shortly after Sarbanes-Oxley was enacted, the Auditing Standards Board produced Statement on Auditing Standards 99 (SAS 99), Consideration of'Fraud in a Financial Statement Audit [Auditing Standards Board, 2002]. As its name suggests, this standard requires auditors to search aggressively for fraud. We will look at SAS 99 later in our discussion of fraud control.

**Basel II**

Banks who do business internationally will also have to comply with the Basel II accord. Basel II requires banks to have capital reserves to cover probable risks. Banks that do not have solid financial reporting systems or risk controls in place must set aside more capital reserves. This reduces the amount of loans they can support, and this in turn limits profits. Basel II gives banks a direct incentive to invest in internal controls to reduce risks.

**Privacy Laws**

Several laws have affected requirements for privacy and for the disclosure of private information. These include the following:

- The European Union (E. U.) Data Protection Directive of 2002 is a broad set of rules ensuring privacy rights in Europe.

- Although the E. U. Data Protection Directive is the most important international privacy rule, many other nations with which U.S. firms do business are also developing strong commercial data privacy laws.

- The U.S. Gramm-Leach-Bliley Act (GLBA) of 1999 requires strong data protection in financial institutions.

- The U.S. Health Information Portability and Accountability Act (HIPAA) of 1996 requires strong protection for private data in health care organizations.

- The U.S. Patriot Act of 2001 gives the U.S. government broad powers to see personal data.

## The Compliance Age

These laws and many others have made compliance a major issue for information systems in all corporations. Unfortunately, IS education has not kept pace with the growing importance of compliance in IT management.

## Control Frameworks

To achieve compliance with SOX and other crucial regulations, companies typically adopt a control framework. Control frameworks help organizations understand all of the things that they need to do and how to do these things.

## Types of Controls

All of these frameworks focus on controls. Earlier, we saw in Figure I that the purpose of controls is to help organizations keep their organizational processes on track to achieve the firm's goals. Figure I also shows that controls generally fall into one of three categories.

- Preventive controls attempt to keep deviations from occurring in the first place. In movie theaters, one person sells tickets but another collects them. This is the segregation of duties. Unless the two parties collude, the person accepting the money for tickets cannot collect money, pocket it, and then allow the person in without giving them a ticket.

- Detective controls attempt to detect deviations when they occur, so that action can be taken. Periodic reconciliations between independent processes will make it likely that deviations in one of the processes will be revealed. In the case of movie theaters, management reconciles the number of tickets sold with the number of tickets collected at the end of each day.

- In some cases, there are corrective controls, which fix deviations. The restoration of backup files on a computer compromised by an attack is a corrective control.

This taxonomy is not perfect. For instance, if people realize that detective controls are in place, this should deter them from misbehavior, and this would be a preventative control. However, the general concept of preventative, detective, and corrective controls is useful in practice.

## Coso

For Sarbanes-Oxley, the PCAOB explicitly requires corporations to use a welldeveloped comprehensive control framework. Although the PCAOB does not require corporations to use a specific framework, it has specifically listed only a single framework as acceptable, and most companies are using this framework to implement SOX. This is the COSO framework.

**The COSO Framework**

Although COSO is universally known by its acronym, the COSO framework actually is a document called Internal Control-Internal Framework [COSO, 1994]. The acronym COSO comes from the organization that created the document, the Committee of Sponsoring Organizations of the Treadway Commission (www.coso.org).

**Objectives**

Control frameworks, as shown in Figure 1, require objectives. In the COSO framework, there are three objectives.

- Operations. The firm wishes to operate effectively and efficiently. It is necessary for the finn to control its general internal operations to do this.

- Financial Reporting. The firm must create accurate financial reports. This, of course, is the focus of Sarbanes-Oxley.

- Compliance. The firm wishes to be in compliance with external regulation. In this paper, we are only directly concerned with SOX compliance.

**Reasonable Assurance**

Good controls cannot completely guarantee that goals will be met. However, an effective control environment will give reasonable assurance that goals will be met.

**COSO Framework Components**

Figure 4 shows the COSO framework. It shows that the framework has five components. These are components rather than phases because there is no time ordering among them. Each must occur simultaneously, and each feeds into others constantly.

**Figure 4: The COSO Framework**

- Control Environment. The component at the base of the COSO framework is the corporation's control environment. This is the company's overall control culture. It includes the "tone at the top" set by top management, the company's commitment to training employees in the importance of control, the punishment of employees (including senior managers) who violate control rules, attention by the board of directors, and other broad matters. If the broad control environment is weak, other control elements are not likely to be effective.

- Risk Assessment. More specifically, a company needs to assess the risks that it faces. Without systematic risk analysis, it is impossible to understand what level of controls to apply to individual assets. Risk assessment must be an ongoing preoccupation for the firm because the risk environment constantly changes.

- Control Activities. An organization will spend most of its control effort on the control activities that actually implement and maintain controls. This includes such things as approvals and authorization, IT security, the separation of duties, and other matters we will look at later. Controls usually have two elements. One is a general policy, which tells what must be done. The other is procedures, which tell how to do it.

- Monitoring. Having controls in place means nothing if organizations do not monitor and enforce them. Monitoring includes both human vigilance and audit trails in information technology. It is essential to have an independent monitoring function that is free to report on problems even if these problems deal with senior management.

- Information and Communication. For the control envirom-nent, risk assessment, control activities, and monitoring to work well, the company needs to ensure that it has the required information and communication across all levels of the corporations.

**Types of Control Activities**

Internal Control-Internal Framework [COSO, 1994] does not list a comprehensive set of control activities, probably because it is impossible to create a complete list of possible controls. However, the document does provide several lists of types of control activities. For instance, on Page 49, the framework notes the existence of manual controls, computer controls, and management controls. On the following page, it provides the following list to consider:

- Top Level Review-comparing budgets with actual performance, tightly monitoring major initiatives, and so forth.

- Direct Functional or Activity Management-managers who run individual operations must examine the appropriate reports for their level, for instance, loan performance in a bank's lending operations.

- Information Processing, including the enforcement of manual procedures, such as checking if a customer's accounts payable are not above a certain amount before accepting an order. Note that information processing must focus on business processes, not merely on IT processes.

- Physical Controls-inventories, locked cash drawers, write-only archival media, and so forth.

- Performance Indicators-Relating different sets of data to each other to check for inconsistencies, noting deviations from normal performance (in either direction), unusual trends, and so forth.

- Segregation of Duties - requiring most sensitive processes to be completed by two or more people, so that no single person can engage in improper activities without this becoming apparent. For instance, one person may purchase and order, but another person will record it. Alternatively, no single person can both authorize and make a purchase.

**Controls for Information Systems**

On Pages 52-55, the document lists some controls over information systems. At a most basic level, the framework discusses the differences between application controls and general controls. Application controls, as the name suggests, involve individual applications (accounting applications, spreadsheets, and so forth), including manual operations in using them. This includes interfaces to other systems for input data, checks on input, internal checks during processing to flag errors and misbehavior, and so forth.

General controls cover everything beneath the applications-computers, operating systems, the network, and so forth, together with manual operations in using them. This includes making purchases, application systems development, maintenance, access controls, evaluating packets software, and so forth. The controls needed in individual applications will depend on the quality of general controls.

**Controls over "Evolving Issues"**

The report spends approximately half of Page 55 on "evolving issues." Two brief paragraphs are devoted to end-user computing (EUC). The first paragraph simply says that EUC exists. The second give the following meager guidance:

To provide needed control for EUC systems, entity-wide policies for system development, maintenance, and operation should be implemented and enforced. Local processing environments should be governed by a level of control activities similar to the more traditional mainframe environment. (COSO, 1994, p. 55)

Internal Control-Integratedftamework does not give any specific guidance on spreadsheets. In fact, it does not mention them. In general, this is an old (1994) document that was written before spreadsheets become important, or, probably more correctly, before IT control professionals realized that spreadsheets were important.

## CobiT

COSO is a general control planning and assessment tool for corporations. For IT controls, there is a more specific framework, CobiT (Control Objectivesfor Information and Related Technologies) (IT Governance Institute, 2000). The IT Governance Institute has not only created the control objectives framework. The institute also has developed detailed guidance for implementing the CobiT framework.

**The CobiT Framework**

Figure 5 illustrates the CobiT framework. This framework has four major domains:

**Figure 5: COSO/CobiT Framework**

- Planning and Organization. The four domains follow the IT life cycle. The planning and organization domain has I I high-level control objectives that cover everything from strategic IP planning and the creation of a corporate information architecture to the planning of specific projects.

- Acquisition and Implementation. After planning, companies need to acquire and implement information systems. This domain has six high-level control objectives.

- Delivery and Support. Most of an IT project's life takes place after it implementation. Consequently, the CobiT framework has 13 high-level control objectives for delivery and support. This is more than any other domain.

- Monitoring. Finally, firms must monitor their processes, assessing the adequacy of internal controls, obtaining independent assurance, and providing for independent audit.

Although the four domains define the scope of CobiT, they are only the beginning of CobiT. Beneath the four domains are 34 high-level control objectives, which Figure 5 also shows. Beneath these, in turn, are more than 300 detailed control objectives. CobiT also includes many documents that help organizations understand how to implement the framework.

**Dominance in the United States**

The IT Governance Institute is the child of the Information Systems Audit and Control Association (ISACA). ISACA, in turn, is the primary professional association for IT audit professionals in the United States. The association's certified information systems auditor (CISA) certification is the dominant certification for U.S. IS auditors, so it is not surprising that CobiT has become the dominant framework for auditing IT controls in the United States.

**COSO and CobiT**

Obviously, both COSO and CobiT pertain to information technology used in financial reporting. Figure 5 shows how CobiT relates to COSO at a broad level. This figure illustrates that it is relatively simple to combine COSO with CobiT at a conceptual level, although the details are anything but simple.

## Other Frameworks

Although COSO and CobiT have dominated Sarbanes-Oxley planning in the United

States, other planning frameworks are also important.

**ISO/1EC 17799**

Figure 6 shows the relationship between COSO, CobiT and three other frameworks-ISO/IEC 17799, Common Criteria, and ITIL. The figure emphasizes that the three frameworks overlap but focus on somewhat different things.

**Figure 6: COSO, CohiT, 17799, Common Criteria, and ITIL**

For example, CobiT, as its name implies, focuses specifically on controlling the entire IT process, while COSO focuses more broadly on financial reporting controls. In contrast, ISO/IEC *17799, Code ofPracticefor* Information Security Practice (ISOJEC, 2000), as its name suggests, focuses more narrowly, on IT security. Security is part of IT controls, of course, so *17799* can help in creating IT controls.

ISO/lEC 17799 grew out of an earlier standards effort by the British Standards Institute. In *1995,* the Institute produced BS *7999.* This standard has two parts. ISO and the IEC adopted Part I as *17799.* This first part is a broad code of practice.

For organizations wishing certification of their standards effort, Part 11 of *7999* (Information Security Management System) has auditable controls. Consequently, many companies have chosen to be compliant with *17799* by being certified in Part 11 of *7999.*

In other frameworks, including COSO and CobiT, companies certify themselves, sometimes with the concurrence of an external auditor. They lack *17799's* third-party certification process, which external parties may value highly.

Even *17799 is* quite complex. Broadly, it has ten major categories' of standards. It subdivides these into *127* specific controls. At the time of this writing, *17799 is* in flux. ISO is preparing to release a new version in mid-2005. Although there has been much speculation about the revisions in the standard, it seems best to hold comments until the final 2005 version. In addition, it seems likely that ISO and IEC will adopt BS *7999* Part 11 as an ISO standard. Again, however, we must wait for ISO and IEC to act.

**Common Criteria**

Figure 6 shows that the Common Criteria (ISO/IEC 15408, *Information technology -- Security techniques -- Evaluation criteriafor IT security)* standard is even more specific, Common Criteria specifically focuses on the evaluation of security products, such as firewalls. It provides a way for purchasers to know specifically what security features a security product claims to offer and how rigorously the product was developed. However, the Common Criteria approach has somewhat limited use because it is difficult to apply and does not provide a high level of assurance that a product actually is secure.

**ITIL**

Another framework for IT is the Information Technology Infrastructure Library (ITIL). ITIL is a broad set of best practices guidelines for providing IT services. It is widely used in Europe and is becoming popular elsewhere. ITIL is highly processoriented, specifying systematic approaches to implementing security and other IT services. ITIL best practices may be helpful in implementing other guidelines. However, ITIL does not have the detailed guidance necessary for developing and implementing IT financial reporting controls.

[1] Security Organization (including relationships with third parties and outsourcers). Security Policies (to provide management direction and support). Asset Classification and Control. System Development and Maintenance (building in security). System Access Control (prevention and detection of unauthorized access). Computer and Network Management. Physical and Environmental Security, Compliance (avoiding breaches of laws). Personnel Security (to reduce human errors, fraud, theft, or unauthorized usage). Business Continuity Planning.

## Spreadsheet-Specific Guidance

If one does an Internet search with the terms "spreadsheet" and "Sarbanes," there will be many hits. Nearly all of these, however, are about spreadsheets used to document SOX compliance, not how to control spreadsheets used in financial reporting.

The one major exception to this silence on spreadsheet control for Sarbanes-Oxley is a six-page report by PriceWaterhouseCoopers [20041. This report lists a large number of controls. The first step is to inventory all of a firm's spreadsheets that are "in scope," that is, are used in financial reporting. Next steps are to evaluate the riskiness of these spreadsheets, determine necessary controls, evaluate the existing (as-is) controls on these spreadsheets, and develop action plans for remediating control deficiencies. Once concern is the report's method for assessing riskiness. It lists nine factors to consider when evaluating the "risk and significance" of a spreadsheet:

- Complexity of the spreadsheet and calculations
- Purpose and use of the spreadsheet
- Number of spreadsheet users
- Type of potential input, logic, and interface errors Size of the spreadsheet
- Degree of understanding and documentation of the spreadsheet requirements by the developer
- Uses of the spreadsheet's output
- Frequency and extent of changes and modifications to the spreadsheet
- Development, developer (and training) and testing of the spreadsheet before it is utilized

Although this is a good list, it appears to be based on an incorrect view of spreadsheet errors. Research indicates that the largest indicator of the number of errors in a spreadsheet is simply the length of the spreadsheet. There will be errors in about 2% of all fon-nula cells if there is no deep testing. This also is the case in programming. However, from programming, we know that more complex programs have more errors than simpler programs, but only by a factor of about four [Zage & Zage, 1993]. Consequently, long simple spreadsheets will have many more errors than short complex programs. Certainly, reporting complexity first and size half way down the list is a concern.

Even more of a concern is placing testing last in the list. As discussed in the following sections, a lack of testing is a devastating control deficiency, not simply one part of a last concern on a list.

The report also lists a number of controls that should be considered to mitigate risks inherent in the spreadsheet environment.

- Change control: the authorization of change requests, testing the spreadsheet, and formal sign-off by another individual.

- Version control: Ensuring that only the current and approved version of each spreadsheet is used. Naming conventions that include version numbers and dates and the use of structured directories can help in this.

- Access control: assign appropriate access rights to people who need to use the spreadsheet. Use a password to control access.

- Input: Whether manual or automatic data entry is used, there should be reconciliations.

- Security and data integrity: storing spreadsheets in protected directories, and locking formula cells to prevent logic changes.

- Documentation: Ensure that the business objective and functions of the spreadsheets are understandable.

- Development life cycle: Use a standard systems development life cycle. The report specifically says that testing is critical (but does not discuss how to do testing).

- Backup and Archiving: spreadsheets should be backed up because of their sensitivity. They should be archived in read-only fonriat for later review.

- Logic inspection by an independent person other than the developer: The report does not discuss how logic inspection is different from testing. Probably, testing is taken as what we call execution testing later in this paper. The report appears to sanction logic checking by a single individual. As discussed later, this appears to be unwise.

- Segregation of duties/roles, and procedures: There should be authorities, roles, and procedures for ownership, sign-off, and other matters. This item is so brief in the report that it provides little guidance.

- Analytics, In accounting, "analytics" refers to calculations built into spreadsheets (ratios, cross-checks, statistical patterns, etc.) that help detect errors and fraud.

The report says that firms should enforce these controls. For instance, changes should be made independently in two copies of each spreadsheet, and the two copies should be compared. In another example, a sample of cells can be tested to ensure that they are password-protected if they should be. In addition, names should include modification dates, and names should be compared with modification dates as recorded by the operating system.

Finally, the report gives a number of suggestions for remediation. Specific responsibilities should be assigned to specific people, remediation efforts should be prioritized, and remediation deadlines should be established.

Although this report offers valuable advice, its error beliefs are inadequately, and its lack of detail leave many questions unanswered. One important concern is testing. We will discuss testing in some detail in this paper. Another issue is how to control manual processes in spreadsheet usage.

# The Testing Control

Although spreadsheets need many controls, we will focus on testing because of testing's extreme importance in software quality assurance.

Although fault rates in programming are very similar to error rates in spreadsheet models, programmers and spreadsheet developers react very differently to these defect rates. Programmers, appalled by their fault rates, spend a great deal of time on programming. In a sample of 84 projects in 27 organizations, Jones [1998] found that the amount of time spent in testing to reduce errors ranged from 27% to 34% depending on program difficulty. In every case, subjects reported that insufficient time was planned for testing (p. 139). In another study, Kimberland [2004] found that Microsoft software development teams spent 40% to 60% of their total working time in testing.

The situation in spreadsheet development is very different. As noted earlier, studies since the earliest days of spreadsheet development have found that few organizations have policies requiring extensive testing and that few spreadsheet developers test their spreadsheets intensively in ways similar to those that programmers use.

## Testing in Software Development

Both field audits [see Figure 3] and experiments [Panko 2005b] have shown that spreadsheet models and software programs are very similar in error rates and error types. Consequently, to get insight into spreadsheet testing, it would be good to understand software testing.

### Execution Testing

There are two general ways to test software for errors. One is to use execution testing, in which the tester assembles a large number of test values with known results and runs these test numbers through the program as input values. We will discuss execution testing again under spreadsheet testing.

### Code Inspection

Another method is code inspection, which Fagin [ 1976] developed. In code inspection, a group of testers independently examines a program line by line looking for errors. They then meet to compare their results and attempt to find new errors. Although variations on this basic pattern exist, all code inspection methods involve groupwork and line-by-line code examination.

One tenet of code inspection methodologies is that testers should report their results. The programming inspection data in Panko [2005a] are extensive because so many code inspection studies have reported their results. This reporting is important because it forces software developers to face error rates directly.

#### Code Inspection Parameters

Fagin [ 1976], based on his personal experiences implementing code inspection, emphasized several principles for code inspection. Subsequent experience has reinforced his insights.

- Module Size. Perhaps the most important thing is to test small modules rather than entire programs. A typical module size is 100 to 200 lines of code. Although there disagreement about the optimal module size, there is no doubt that keeping modules small is important. For instance, Bernard and Price [1994] found that code inspectors found 72% more errors when modules were shorter. If the module is too long, the tester will not be able to maintain his or her focus during testing.

- Inspection Rates. Another key principle is that inspections should not be rushed. Inspection rates often are measured as lines of code inspected per hour. Basili and Pericone [ 1993 ] found that when the inspection rate rose from 50 lines per hour to 200 lines per hour, the detected fault rate fell from 1.6% to only 0.6%. This was not because programs had fewer faults but because faster inspection detected fewer errors. Russell [ 1991 ] found an even sharper drop in detection when the inspection rate rose from 150 lines per hour to 750 lines per hour. Ebenau and Strauss [ 1994] found that "hasty" code inspections found errors in 1.3% of the lines of code they examined, while non-hasty code inspection found errors in 2.0% of all lines of code. Weller [ 1993 ] also found sharp drops in detection rates as inspection rates increased. Levy and Begin [ 1984], in research on proofreading, similarly found that faster proofreading speed substantially reduced the percentage of spelling errors detected.

- Formula Inspection Speed. One thing that has not been studied in the code inspection literature but that probably is important is the need to slow down even further for complex equations. In proofreading, Healey [ 1980] looked at error detection for nonword errors (errors that do create a dictionary word) in prose for varying word lengths. Detection rates were over 90% for words of 2, 3 or 4 letters. For longer words, however, detection rates were only about 75%. Panko [ 1999], in a study of spreadsheet code inspection, found that error detection rates for long formulas were much lower than for shorter formulas. Just as drivers should slow down for hazardous road conditions but rarely do [e.g., Howarth, 1988; Svenson, 1977], code inspectors need to force themselves to slow down for long formulas.

- Team Size. Another tenet of code inspection is that code inspection must be done by groups rather than by single inspectors. Weller [ 1993 1, who analyzed more than6,000 code inspection studies, found that teams of four found 29% to 108% more errors than teams of three depending on the inspection rate. This limited ability of individuals to find errors also been seen in laboratory code inspection experiments. In these experiments, subjects found only 22% [Johnson & Tjahjono, 1997], a third [Myers, 1978] and a half [Basili & Selby, 1986] of the programming faults seeded in a test program. When Johnson and Tjahjono [ 1997] had subjects work in groups of three, detection rate doubled. In two spreadsheet code inspection studies, Galletta et al. [ 1992, 1997] found that individual detection rates for a spreadsheet seeded with errors were only 5 1 % and 66% in the two studies. In another study using a similar spreadsheet [Panko, 1999], individuals only found 63% of all errors, while groups of three found 83%. Although this seems like a modest increase, this increase came precisely in the errors that individuals were least likely to detect. Steiner [ 197 1 ] developed a theory of why groupwork is effective. He noted that groupwork is more important for difficult problems than for easy problems.

- Team Composition. It is sometimes said that people who develop code should not be on the inspection team because people are not good at finding their own errors. However, proofreading research suggests differently. Daneman and Stainton [ 1993 found that when proofreading took place twenty minutes after writing, individuals found only 59% of their own word errors but found 76% of word errors made by others. However, when proofreading took place two weeks later, people had higher success reading their own work than did others (83% rather than 68%). If inspection takes place some time after development, individuals should be able to be effective at code inspecting their own work, although this hypothesis has not been tested in programming or spreadsheet development.

- Reasonable Expectations for Detection Yields. The reason for the use of team code inspection is that humans are only moderately good at finding errors. This has been seen in many areas of human cognition, such as proofreading [Panko 2005a]. Team inspection can correct more errors, but even team inspection is far from perfect. Jones [ 1988], based on extensive experience, estimated that each formal code inspection only detects about 60% of all errors (p. 199). McCormick [1983] found that code inspections he studied found 86% of all errors, based on the number of errors found later. Boehm and Basili [2001 ], based on their experiences with many studies, reported that peer reviews only caught a median of 60% of all errors and that this yield ranged from 3 1 % to 89%. At a CeBASE [2001 ] workshop, participants shared data from several studies, collectively indicating that 60% or more defects usually are found in inspections. Due to limited error detection rates, programming projects have to have several rounds of inspection. Studies of late inspections after multiple rounds of inspections still find errors in 0. 1% to 0.3 % of all lines of code [Putnam & Myers, 1992]. Given that most studies of initial code inspection find errors in 2% of all lines of code, this data suggests that even multiple rounds of code inspection only find about 90% of all errors.

- Improvement Takes Time. It would be nice if companies could learn to do code inspection quickly. However, experience has shown that improvement is likely totake two or more years [Haley 1996, Spencer 19931. In general, it takes considerable time for people to become competent in complex skills. For instance, it takes about two years of driving before people cease being menaces on the highway, [Bereiter & Scardamalia, 1993]. In addition, once people are skilled, it may be a considerable period of time before the organization becomes competent using new methods.

## Testing Spreadsheets

From the information just presented, it is clear that spreadsheet errors are a serious threat in all large spreadsheets. In software development, where fault rates are similar to error rates in spreadsheet development, most companies mandate a very disciplined development process. This process involves several well-controlled stages, including requirements definition, design, coding, and testing. All of these stages are important, but the biggest reductions in error rates in programming come from testing. This includes both design reviews and coding tests, but we will focus on coding tests. As noted earlier, approximately 30% of all software project effort is devoted to testing.

### Whiskey Cures

Ineffective home remedies sometimes are called "whiskey cures," after the old aphorism, "Of all the things that do not cure the common cold, whiskey is the most popular." With a whiskey cure, you take a step to reduce harm, but the step is largely or entirely ineffective. One commonly seen activity in spreadsheet development is looking over a spreadsheet's results for reasonableness [Hendry & Green, 1994; Nardi & Miller, 199 1 ]. If a few errors are found, the seeker feels that he or she is very effective at finding errors. If no errors are found, then the seeker feels that the spreadsheet is free of error. Unfortunately, given the difficulty of finding errors noted above even when full cell- by-cell code inspection is used, merely looking over a spreadsheet's numbers for reasonableness is a whiskey cure. In addition, studies [Klein, 1997; Rickets, 19901 have shown that people are not very good at finding errors when they assess numbers for correctness. Although looking over results for reasonableness is simple and inexpensive and should be done, it must not be considered an acceptable stopping point.

Another whiskey cure is depending upon developers to test their own spreadsheets during development. However, testing during development is **neither** new nor very effective. When Allwood [ 1984] watched students working on algebra problems, he noted that they frequently stopped to go back and check their work. In some cases, they **did this** when they thought they had made an error. In other cases, they performed what Allwood called a standard check, going back over their work on general principles. Hayes and Flower [ 1980] saw that when people write they constantly shift among planning, writing, and reviewing activities. Reviewing is like the standard check in Allwood's study. Kellog [ 1994] found that people spend 20% to 25% of their time of their writing time reviewing, In an earlier study cited in his 1994 paper, Kellog found that sbjects spent 19% of their time reviewing and also found that reviewing increased as writing time increased.

In spreadsheeting, Panko and Halverson [ 1997] also watched many of their individual subjects and groups during spreadsheet development. They also noted frequent error checking during development but only rare post-development testing. However, many errors remained undetected despite this type of review during development.

**Execution Testing**

A common form of software testing is execution testing, in which the tester inputs values that have known output and observes whether the program gives the correct results. This approach seems obvious, but it often is very difficult to do well. In many cases, the spreadsheet is doing calculations that have never been done before. Consequently, finding the correct output values for test cases may not be feasible or may be prohibitively difficult.

In addition, determining proper test cases is a subtle skill. Most people tend to pick test cases that will prove a program to be correct. This is in line with the human confirmation bias [Snyder & Campbell 1980]. Proper execution testing, however, requires the tester to select cases that are likely to break the program. This requires a great deal of training and experience. Unfortunately, spreadsheet developers rarely have the proper training and experience.

Overall, execution testing in spreadsheet development is likely to be most effective for testing *changes* to spreadsheets. After a spreadsheet is developed and tested, test cases should be developed, and the actual values should be taken as correct values. If the spreadsheet is modified afterwards, the test cases should be rerun. Variables that should not change due to the modifications should be the same as the original values. If they are not, this will indicate an error.

**Code Inspection**

Although spreadsheet developers are not likely to be good at developing test cases for execution testing, they are likely to be quite good at code inspection. They usually are subject experts in the content domain of the spreadsheet. This means that they are likely to understand both the formulas in the spreadsheet and the flow of logic among the formula cells.

Given the threat analysis presented earlier and the discussion of software code inspection presented earlier, spreadsheet code inspection should have the following characteristics. This is not a complete list, but it emphasizes the points that are likely to be the most difficult to implement.

- Inspect All Cells. There is no excuse to say that code inspecting all cells is impractical. Not doing it guarantees an unacceptable error rate. In addition, if programmers can do it, spreadsheet developers can do it. Unless code inspection inspects all cells, it is a whiskey cure.

- Use Group Inspection. A single person will be unable to find many errors in a spreadsheet. The only question is how large the group should be. As noted earlier, Panko [ 1999] found that the detection rate rose from 63% to 83% when individuals gave way to groups of three. If this finding holds, then larger groups will be desirable.

- Inspect Modules. Trying to inspect an entire spreadsheet is a guaranteed route to failure. Inspections should be conducted on each module after it is developed. The optimum module size is unknown, but in Panko's [ 1999] experiment, a spreadsheet with only about 20 unique formulas took students approximately 45 minutes to inspect. A maximum inspection time for a module is about two hours if attention is to be preserved, so optimum module sizes are likely to be fairly small. Most spreadsheets will have many modules and so will require many inspections. The optimum size in terms of cost and detection rates needs to be determined empirically.

- Slow down for complex formulas. As noted earlier, Panko's [1999] study showed that inspection success falls for cells that contain longer formulas. It seems likely that when inspectors reach longer formulas, they should slow down their testing rate dramatically.

- Multiple Rounds of Inspection. If program inspection is a good indicator, each round of inspection is only likely to detect 60% to 80% of all errors, and achieving even these levels is likely to require considerable experience.

## Actual Testing

How do real-world companies do testing? As noted earlier, when the first author asked financial professionals and officers whether their firms did testing for material spreadsheet, most did not, but a small subset said that they did. However, the next question asked respondents how they actually tested their spreadsheets of material importance in financial reporting. As Figure 7 shows, 73% of the respondents said that when their firms test spreadsheets of material importance, they test only some cells. In other words, they consider "looking the spreadsheet over" to be testing. Only 12% said that their firm tested all cells in the spreadsheet, and only 2% said that their firm used both multiple testers and cell-by-cell testing.

**Figure 7: Testing Methods for Material Financial Spreadsheets**

### Spreadsheet Auditing Programs

One possible approach to testing is to use spreadsheet auditing programs that are designed to find errors in spreadsheet programs automatically. Obviously, using a spreadsheet auditing program is faster and less expensive than manual code inspection.

Unfortunately, to borrow a term used in U.S. prescription drug regulations, spreadsheet auditing programs have not been proven to be "safe and effective." No research known to the authors has established what percentage of errors auditing programs will find. Unless this percentage is very high, it will not be safe to use spreadsheet auditing programs in lieu of full manual code inspection.

To give an analogy, spelling checkers in word processing programs will catch all spelling errors in which the mistyped word is not a dictionary word. However, if the word is in the dictionary, the spelling checker will ignore it completely. Proofreading research has shown that the foriner type of error is, also easiest type of spelling error for humans to detect. Grammar checkers, in turn, are frustrating for missing grammar errors and for calling correct grammar incorrect.

We suspect that auditing programs will be even less precise than grammar checkers. They are unlikely to be able to find omission errors, such as if the developer leaves out paying the monthly rent. Nor are they likely to be able to catch most logical formula errors based on the use of the wrong algorithm or the incorrect expression of an algorithm. Errors in typing numbers are also likely to escape these programs. It seems like they would be best in pointing errors, but these are quite uncommon [Panko & Halverson, 1997, 2000]. In programming, Endress [ 1975] found that 48% of all errors are problem misunderstandings. In spreadsheet development, auditing programs are not likely to be effective with such errors.

In general, because there are multiple types of errors, almost any error reduction or detection method is likely to be limited. Beizer's [1990] Pesticide Paradox reflects this problem of multiplicity in dealing with errors in the context of programming: "Every method you use to prevent or find bugs leaves a residue of subtler bugs against which those methods are ineffective."

In marketing materials, spreadsheet auditing programs focus on the ability to detect patterns in how blocks of cells are laid out and on certain types of link geometries. Panko [2000] has listed the most common errors for two spreadsheet development tasks. Few of these errors seem amenable to detection based on geometry.

The most common advice to writers is to use automated spelling and grammar checking functions for "pre-proofing" of a document before full manual proofreading is done. The same may be the case with automated spreadsheet auditing programs.

Fortunately, Excel has a number of pre-auditing tools that can be used by everyone. Excel tags many questionable formulas with green comment markers. By going to Tools, Error Checking, the user can step through these potential errors in order. One choice usually is Evaluate Formula, which steps the auditor through a formula's calculation term by term.

In addition, for manual code inspection, a user can see the precedents of a formula (cells to which a formula refers) by double-clicking on a formula cell or clicking on the cell and hitting F2. Cell references and the cells or ranges to which they refer are shown in the same color.

By bringing up the formula auditing toolbar (View, Tool Bars, Formula Auditing) the user will have several options, including showing subsequent formulas that refer to a particular formula (decedents). A cell with no decedents has to be a bottom-line value. Any cell in a spreadsheet that is not a bottom-line value yet has no decedents is suspicious.

# Other Error Controls

Although we have focused on testing as a sine qua non for reducing errors, other controls are needed to achieve low error rates.

## Design and Design Reviews

In programming, the standard systems development life cycle calls for strong initial planning, in which the systems analyst determines needs, in which a programmer creates a design for the program, and in which the organization conducts a design review. Errors found during design reviews are much less expensive to fix than errors in later phases, especially coding. Studies [McCormick, 1983; Jones, 1986] have found that more errors occur during the requirements and design stages than during the coding phase, so design reviews should be able to reduce error reduction costs considerably.

Unfortunately, not only is design review rare in spreadsheet development but design itself tends to be limited. In experiments, observations of people developing reasonably challenging spreadsheets has shown that almost all subjects begin entering labels, numbers, and formulas within a minute or two at most [Brown & Gould 1987, Panko & Halverson 1997, Takaki 2005].

Given the cost effectiveness of design reviews and the danger of creating difficult-to-understand spreadsheets if there is insufficient planning, companies need to develop mandatory design and design review processes for all nontrivial spreadsheets.

At the same time, when people develop spreadsheets, they may not understand the situation fully. The process of building the spreadsheet model itself may be the best way to understand the situation [Panko 1988] because the developer can work until hitting a snag that must be addressed or because the developer may begin to develop a spreadsheet and realize that the initial design will not work. In addition, as the user begins to do calculations, he or she might find surprising results. These results can lead to new directions in design. Similarly, in writing, Kellog [ 1994] noted that inspiration generates production, which generates more inspiration.

Of course, spreadsheets that grow "organically" may have many design and construction flaws. One possibility is to use such spreadsheets as prototypes and then to build production systems from scratch, using the prototype's general design. However, this takes a great deal of discipline. In writing, Flower and Hayes [ 1980] have suggested that writers should have "plans that are detailed enough to test, but cheap enough to throw away." The same should be true for spreadsheet developers.

No matter how the overall spreadsheet is designed, it seems important to design individual formulas before constructing them. The first author's observations of laboratory subjects developing spreadsheets indicated that they often entered a formula they were not sure of and then looked at the results to see if they looked right. This "ready, fire, aim" approach is fast, but it is not safe.

# Formula Protection

As noted earlier, users sometimes "hardwire" spreadsheets by typing numbers into formula cells. Dent [ 1994] reported that hardwiring was very common in a mining firm. In another case, Rabbit Photo in Australia uses a spreadsheet program for budget and expense consolidation. The company found that users were constantly overwriting formula cells to enter "correct" values. When the company turned on cell protection to prevent the overwriting of formulas, users turned off protection and made their changes anyway. The company eventually moved to a centralized software package with a spreadsheet-like interface. Users could not overwrite formulas in this program. Today, companies can use passwords to prevent users from turning off cell protection, but this will require the selection of good passwords. Given the wide extent of the hardwiring problem, companies need to address cell protection actively.

## Readability

In a badly constructed spreadsheet, readability will be so poor that testing will be extremely difficult. It is important to implement controls to ensure readability.

- Modules. Testing requires reasonable-sized modules. Consequently, it is important to develop spreadsheet models as a group of modules. This can also improve readability.

- Flow. In most countries, we read from left to right and then from top to bottom. It is important for readability to have this type of clear basic flow in the spreadsheet's logic. Boehm and Basili [20011 found that good architecture in programming can reduce error correction time dramatically. In spreadsheets, Saariluoma [ 1989] found that poor logic flow could dramatically increase the time it took to understand spreadsheets.

- Short Arcs of Precedence. Raffensperger (2003) has given a number of suggestions to improve spreadsheet readability. One is to use short arcs of precedence between each formula and the cells that precede it. This requires formatting to be simple so that there is not too much vertical white space. Keeping arcs of precedence also is consistent with spreadsheet error research. Lerch [ 1988] found that people often make errors when entering formulas that refer to distant cells. Janvrin and Morrison [1996, 2000], in turn, found that the error rate was extremely high when links had to be made to data or formulas in other worksheets in a spreadsheet workbook.

- Panko [1988] suggested two basic principles for entering formulas. One was to avoid jamming, in which a cell formula contains multiple numbers. For instance, if a company sells SOO units of a product at $400, the developer might enter a cell with the content "=500*400. This is jamming. It is better to have one cell labeled units and containing the number 500 and a second cell labeled price with the value $400. The revenue cell, then, would refer to the other two cells. In short, there should be a single number in a cell or no numbers at all. Teo and Tan [ 1997] confirmed that jamming leads to increased errors during subsequent what-if analysis. With jamming, it is difficult to know which cells to change because individual numbers do not have text labels to identify them. Avoiding jamming also improves readability because readers can understand the labeled numbers more easily. For this same reason, avoiding jamming also makes testing easier. Jamming does have one advantage: it shortens arcs of precedence. However, this is a small benefit compared to jamming's loss of readability.

- Panko [ 1988] also suggested that numbers should not be entered more than once. For instance, if units sold is used in both revenue and cost calculations, the number of units sold should be entered into a cell only once and then referred to in revenue and cost fonnula cells. Lukasic [ 1998] examined two large spreadsheets and found that the same number was often entered multiple times and that the number was not always changed consistently in what-if analyses.

- Repeat numbers across multiple columns through referencing. Teo and Tan [ 1997] did find one situation in which repetition is valuable. In a time-period-based spreadsheet, when students entered numbers like tax rates in all columns of a spreadsheet, they made fewer errors during what-if analysis than if only a single cell had the tax rate. This makes sense because Lerch [1988] found that when people entered references to cells in different rows and columns they made many more mistakes than they did when referencing cells in the same row or the same column as the formula cell. Even so, the actual number should only be typed once and then referenced in other cells in the row that contains it.

### Appropriate Documentation

Almost all spreadsheets have known limitations. There appears to be a need to have a documentation worksheet that gives readers the context of the spreadsheet's development. For instance, the documentation should warn that certain information had to be estimated because time pressures prevented the collection of data. In addition, if input values only have limited ranges over which the spreadsheet has been tested, this should be noted as well.

## Controversies

- The controls just noted all have a foundation in research findings. Several other prescriptions are often given. Unfortunately, they either have no research basis or appear to run counter to research findings.

- Use an Input/Processing/Output Design? A very common prescription is for each module to have an input section containing data, a processing section showing intermediate formulas, and an output section containing the results. There is no research supporting this approach at all. In addition, this approach creates extremely long arcs of precedence, which are very likely to lead to pointing errors when entering formulas. This approach is also likely to harm readability. To understand calculations, the reader would have to switch constantly between the three sections. Note that most tax preparation programs for individuals show calculations with the data the calculations use. This provides high readability.

- Use Long Formulas? Although Raffensperger [2003] has made many valuable suggestions for improving readability, one of his suggestions appears to run counter to research. This is his suggestion to use complex formulas rather than a series of several simpler formulas. As noted earlier, proofreading research shows that people's error detection rate for spelling errors falls rapidly with word size. Panko [ 1999] found that this was also true for spreadsheet formulas.

- Clarity Rather than "Efficiency." It is important to make logic flows as clear as possible, even if there is a shorter way to express logic. Given the speed of today's computer, there is no excuse for "tricky coding." Readability of logic flows is paramount in avoiding errors during development as well as in enhancing testing.

- Use Range Names? Another common suggestion is that developers should use range names in their formulas. Although this appears to be attractive, it raises a serious concern. In selecting ranges for range names, pointing errors can assign the wrong range to a range name. Although the range will be wrong, but it will appear to be correct in formulas that merely reference the range names. Although the research findings are not clear on this issue, using range names should be considered potentially dangerous until research on using range names is done.

- Appropriate Code Documentation. Almost everyone suggests that documentation is important, but it is rarely done in practice because the task is so onerous. One potential option is use code documentation only where the flow of logic is likely to be non-obvious. This could be done with comments inserted into formula cells. This approach might reduce code documentation sufficiently to increase its use. Research has not been done in this area.

- Risk Analysis. A number of prescriptions now call for risk analysis, in which the likelihood of errors and their likely impacts are assessed. Riskier spreadsheets should have more controls. Although the need for risk analysis is obvious, most methodologies have poor approaches to assessing the likelihood of errors.

- Spreadsheet **error research uniformly indicates** that the main factor leading to the likelihood of error is the length of the spreadsheet in terms of calculation chains. Methods tend to focus too much on formula complexity and other factors that are likely to be secondary. It seems mandatory to assume a 2% cell error rate in any risk analysis and to assume, consistent with software research, that 10% to 20% of all errors will be serious.

# Controlling for Fraud

Traditionally, spreadsheet error researchers have focused on "honest error," that is, the assumption that incorrect spreadsheets result from honest mistakes. However, people who commit fraud may use spreadsheets to execute their frauds. Controlling for fraud in spreadsheet development requires new approaches to controlling spreadsheet usage in the firm. Given the importance of Sarbanes-Oxley, controlling for fraud involving spreadsheets has become a serious concern.

# Fraud

We discussed fraud informally earlier. Legally, a fraud exists when one person knowingly lies or conceals information whose nondisclosure would make other statements misleading, in order to get the victim to act in a way contrary to the victim's interests. Note that two elements are needed for there to be fraud: deception and harm. As noted earlier, when people develop spreadsheets, there often is bias that does not rise to the level of fraud. However, when the impact becomes large, then fraud occurs.

# The Fraud Triangle

As noted earlier, the Auditing Standards Board's SAS 99 requires auditors to be extremely vigilant about fraud. Compared to SAS 82, which the new standard replaced, SAS 99 provides much more guidance on the detection of fraud, and SAS 99 requires many more mandatory actions to seek out fraud. SAS 99 requires significant changes in mandatory auditing procedures and documentation in a financial statement audits.

The standard requires a mindset that constantly asks questions with a critical mind in order to identify risks and actual fraud. The standard requires the auditor to look for political promises that will be difficult to meet legitimately, budget pressures, general management style, the effectiveness of the firm's audit committee, the vigilance of the board of directors, and many other things.

One contribution of SAS 99 is the fraud triangle, which is shown as Figure 8. The fraud triangle emphasizes that auditors need to look for three indications of fraud risk.

Figure 8: **The Fraud Triangle**

- Incentive. Auditors must constantly ask why employees would be attracted to fraud in particular circumstance. An obvious motive is the misappropriation of assets, but other motivations are very common. For instance, in many cases, the goal of the fraud is to cover poor performance (as was the case at Allfirst) or to make a firm look like it is doing better than it actually is to attract more capital.

- Opportunity. Even if incentives are strong, if there is little opportunity to succeed, then the existence of strong controls will deter most fraud. Conversely, if controls are weak, this will offer a strong attraction to employees with potentially criminal motives.

- Rationalization. Law enforcement officials have long known that it is important to understand how criminals justify their behavior to themselves. For instance, when the fraud is done to cover up poor performance, **the perpetrator** is likely to believe that they will "make things good" in the future. In other cases, there is a rationalization that "everybody does it," or "the company owes it to me." A strong control environment with intensive ethics training can help reduce rationalizations. It is very important for auditors to understand these motivations and opportunities so that they can accurately assess the potential for fraud in given situations.

## Fraud Controls

Fraud controls are ways to reduce the frequency, severity, or difficulty of detecting frauds. In general, there are four general categories of controls: management controls, operational controls, technological controls, and data controls.

### Managerial Controls

Management controls generally involve the setting and oversight of broad policies, rather than day-to-day actions.

- Top Management Support. At Allfirst and AIB, top management was largely divorced from control issues, leaving them to back-office groups. There was even confusion over whether various control measures should be handled by AlB or Allfirst.

- Risk Assessment. Top management must conduct risk assessments to identify areas of high risk. Certainly, trading large options is highly risky if not properly controlled. In addition, it requires specialized knowledge, and allowing an inexperienced trader to do it in a company like Allfirst, where it was outside of the firm's main line of business, was obviously a high-risk situation that needed extensive control.

- Policy Development. A company needs to establish policies for activities involving high risks. Policies give strong direction to lower-level employees, and they ensure (if followed) that all activities will be treated equally.

- Policy Follow-Up. Creating policies good, but if they are not followed, then they mean nothing. Management must check that its policies are being followed through some sort of auditing function.

- Establishing Independent Audits and Reports. Sarbanes-Oxley requires firms to create audit committees that are independent of the company's senior management. They also require that companies set up a "whistle blower's" hotline for any employee to contact the independent audit committee. This committee's and hot line's scope should extend beyond financial reporting.

- Enforcing Sanctions. When employees violate policies, the firm needs to implement sanctions. Otherwise, employees have only limited incentives to follow corporate policies.

**Operational Controls**

- Good Development Practice. There should be controls on application development for all spreadsheets. This should include initial risk assessment, development methods and practices, and testing.

- Cell Protection. One development practice is particularly important. This is creating cell protection to prevent users from overwriting formula cells. This is always a concern for innocent error, and it is even more of a concern for fraud.

- Operational Audits. Creating policies is good, but if they are not followed, then they mean nothing. Management must check that its policies are being followed through an active auditing function.

- Separation of Duties. There should be clear separation of duties so that at least collusion will be necessary for fraud to succeed. For instance, it may make sense to separate the functions of spreadsheet developer and spreadsheet user.

- Controlling Developer Access. In programming, it is common to have development be done on a single machine to which the developer has access. After development, the program should be moved to a testing machine. On the testing machine, testers should have access, but the developer does not. Finally, the program is moved to a production machine on which neither the developer nor testers have access.

- Requiring Testing. There should be required procedures for testing that specify all of the aspects of testing described earlier.

- Training in Paranoid Testing. Code inspection normally looks for innocent errors. Companies need to develop practices to identify fraudulent deception, including attempts to hide these subterfuges.

- General Employee Training. It is important to train all employees to be wise consumers of spreadsheet results. This includes creating healthy skepticism about the trustworthiness of spreadsheets in terms of both errors and deception.

**Technological Controls**

When firms look at spreadsheets, they usually are dismayed by the poor fraud controls built into spreadsheet programs. This is an area that badly needs to be developed by Microsoft and other vendors. The following is a list of possible technological controls that need to be considered.

- Change Auditing. With spreadsheets today, it is difficult or impossible to know who made changes to the spreadsheet, what they did, and what the spreadsheet said before. Change auditing needs to be added to spreadsheet programs if fraud is to be controllable.

- Usage Auditing. Similarly, when a spreadsheet is used, multiple people may enter data. For the separation of responsibilities during data entry, the spreadsheet program should be able to record who enters which values.

- Authentication. Auditing requires strong authentication. Passwords have proven to be relatively easy to guess or steal. For high-risk spreadsheets, stronger authentication methods should be offered by spreadsheet programs.
- Compilation Tools. One useful feature in spreadsheets would be the ability to compile spreadsheets. Users would only be able to enter data and would not be able to change the logic. Although cell protection should be able to do this, it is too easy to circumvent.

- Audit Detection Tools. In security, when certain audit events occur, intrusion detection systems send warnings to security administrators. Spreadsheet programs need to develop the ability for certain events to be audited during spreadsheet development and use and for warnings to be sent to security administrators when certain events occur.

### Data Controls

Data controls focus on how data is handled. Sources should be controlled, each input item should be checked for reasonableness (length, range of valid values, etc.), data input operators should be trained, and so forth.

# Conclusion

Although Sarbanes-Oxley has focused attention on spreadsheets in financial reporting, once organizations begin to take a hard look at their spreadsheets, they will almost certainly realize that all large and important spreadsheets need careful control.

We know the most about innocent spreadsheet errors. We know that people make errors in approximately 2% of all unique formula cells even with careful development. We know, as a consequence, that all large spreadsheets will contain many errors after careful development. We know that testing is needed to reduce this error rate to a reasonable level, and we generally know how to test spreadsheets and how to develop them so that they are testable.

However, financial reporting is also concerned with fraud. Unfortunately, spreadsheet programs today offer few useful features for dealing with fraud. Financial reporting systems that use spreadsheet programs almost invariably have many manual operations, and spreadsheet programs offer no tools for event auditing. Until spreadsheet programs develop features to address fraud reduction concerns, organizations will have to cope by developing strong management and operational controls.

# Figures

**Figure 1: Controls**

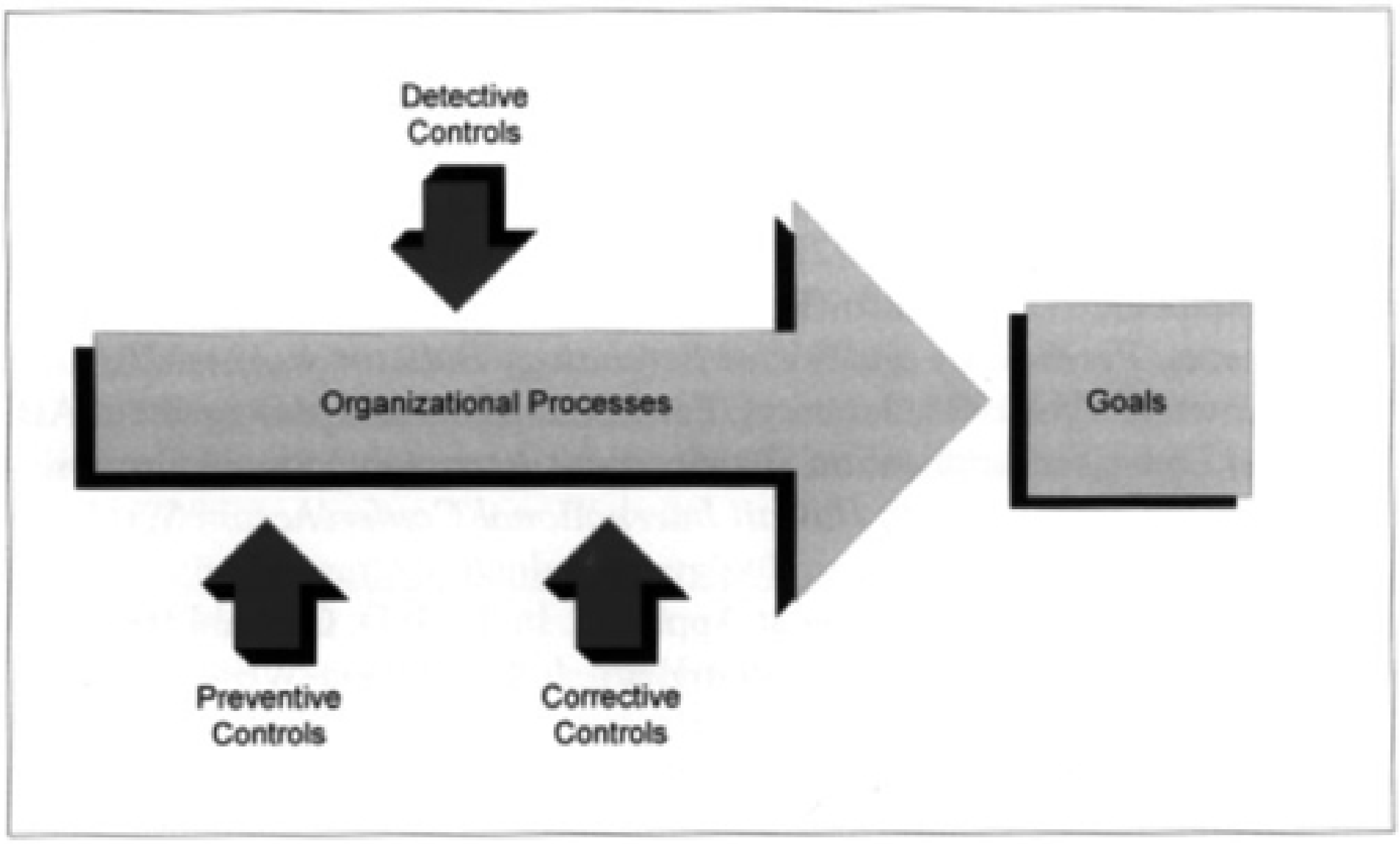

**Figure 2: Testing for Material Financial Spreadsheets**

For spreadsheets of material importance used in financial reporting, what percentage does your company test?

| | |
|---|---|
| Almost None | 24% |
| Under 10% | 20% |
| 11% to 25% | 12% |
| Over 25% | 17% |
| Nearly All | 16% |
| Not Applicable | 11% |
| Total | 100% |
| Respondents | 862 |

**Figure 3: Audits of Real-World Spreadsheets**

| Authors | Year | Number of SSs Audited | Average Size (Cells) | Percent of SSs with Errors | Cell Error Rate | Comment |
|---|---|---|---|---|---|---|
| Hicks | 1995 | 1 | 3,856 | 100% | 1.2% | One omission error would have caused an error of more than a billion dollars. |
| Coopers & Lybrand | 1997 | 23 | More than 150 rows | 91% | | Off by at least 5%. This amount could indicate |
| KPMG | 1998 | 22 | | 91% | | Only significant errors that could affect decisions. |
| Lukasic | 1998 | 2 | 2,270 & 7,027 | 100% | 2.2%, 2.5% | In Model 2, the investment's value was overstated by 16%. Quite serious. |
| Butler | 2000 | 7 | | 86% | 0.4%** | Only errors large enough to require additional tax payments.** |
| Clermont, Hanin, & Mittermeier | 2002 | 3 | | 100% | 1.3%, 6.7%, 0.1% | Computed on the basis of non-empty cells. |
| Interview I* | 2003 | ~36 / yr | | 100% | | Approximately 5% had *extremely* serious errors. |
| Interview II* | 2003 | ~36 / yr | | 100% | | Approximately 5% had *extremely* serious errors. |
| Lawrence and Lee | 2004 | 30 | 2,182 unique formulas | 100% | 6.9% | 30 most financially significant SSs audited by Mercer Finance & Risk Consulting in previous year. |
| **Total** | | **88** | | **94%** | **5.2%** | |

*In 2003, the first author spoke independently with experienced spreadsheet auditors in two different companies in the United Kingdom, where certain spreadsheets must be audited by law. Each audited about three dozen spreadsheets per year. Both said that they had never seen a major spreadsheet that was free of errors. Both also indicated that about five percent of the spreadsheets they audited have very serious errors that would have had major ramifications had they not been caught. Audits were done by single auditors, so from the research on spreadsheet and software auditing, it is likely that half or few of the errors had been caught. In addition, virtually all of the spreadsheets had standard formats required for their specific legal purposes, so error rates may have been lower than they would be for purpose-built spreadsheet designs.

**The low cell error rate probably reflects the fact that the methodology did not inspect all formulas in the spreadsheet but focused on higher-risk formulas. However, error has a strong random component, so not checking all formulas is likely to miss many errors.

Source: Panko (2005)

**Figure 4: The COSO Framework**

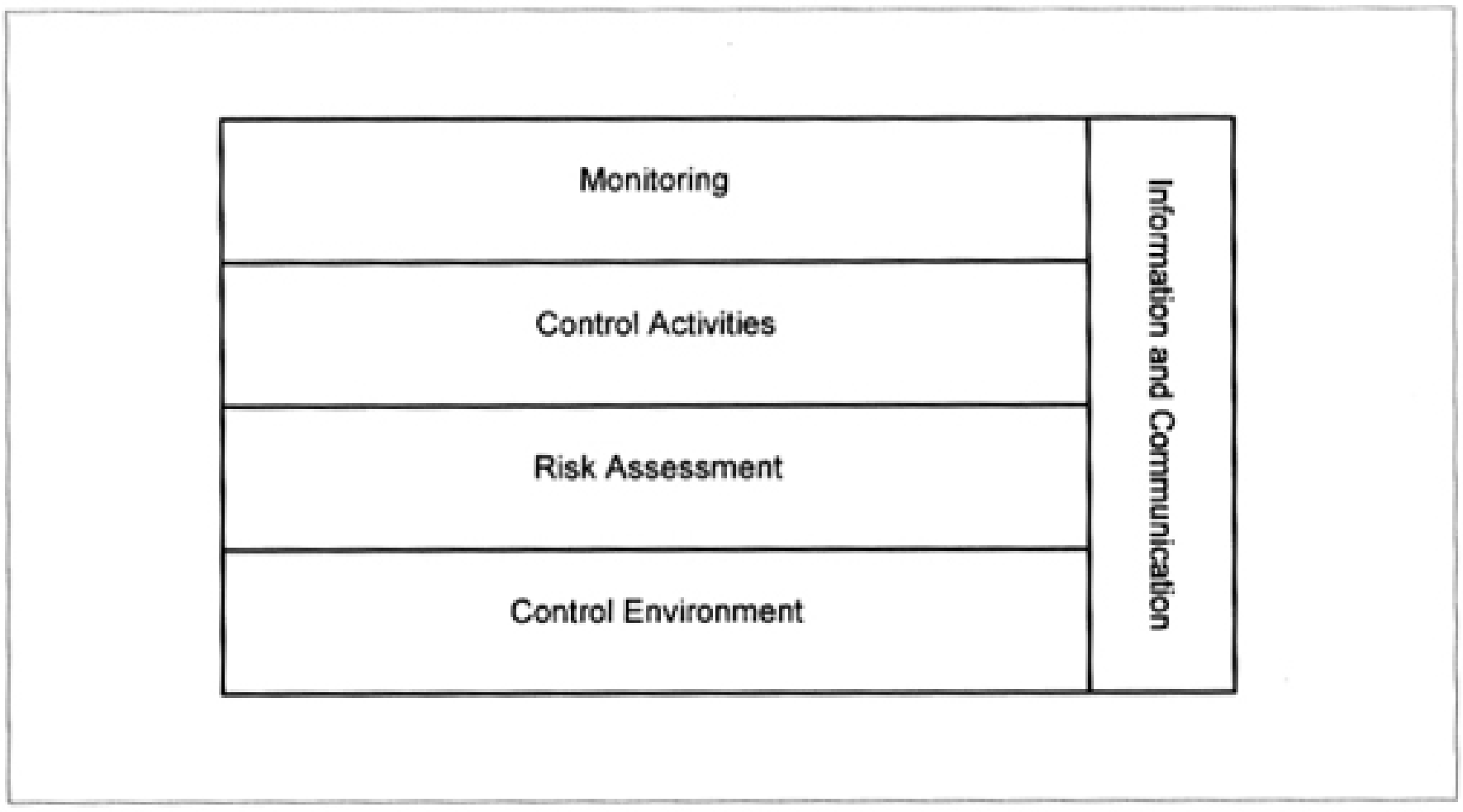

**Figure 5: COSO/CobiT Framework**

| | | | COSO Components | | | | |
|---|---|---|---|---|---|---|---|
| Corporate Level | Activity Level | CobiT Objectives | Control Environ. | Risk Assess. | Control Activities | Info & Comm | Monitoring |
| | | **Planning and Organization** | | | | | |
| X | | IT strategic plan | X | X | | X | X |
| X | | Information architecture | | | X | X | X |
| | | Technological direction | | | | | |
| X | | IT organization/relationships | X | | | X | |
| | | Manage IT investment | | | | | |
| X | | Communication aims & directions | X | | | X | X |
| X | | Manage human resources | X | | | X | |
| X | | Ensure compliance | | | | X | X |
| X | | Assess risks | | X | | | |
| | | Manage projects | | | | | |
| X | | Manage quality | X | | X | X | X |
| | | **Acquisition and Implementation** | | | | | |
| | | Identify automated solutions | | | | | |
| | X | Acquire/Develop app. software | | | X | | |
| | X | Acquire technological infrastructure | | | X | | |
| | X | Develop & maintain procedures | | | X | X | |
| | X | Install and test systems | | | X | | |
| | X | Manage changes | | | X | | X |
| | | **Delivery and Support** | | | | | |
| | X | Define and manage service levels | X | | X | | X |
| | X | Manage third-party services | X | X | X | | X |
| X | | Manage performance and capacity | | | X | | X |
| | | Ensure continuous service | | | | | X |
| | X | Ensure systems security | | | X | X | X |
| | | Identify and allocate costs | | | | | |
| X | | Educate and train users | X | | | X | |
| | | Assist and advise customers | | | | | |
| | X | Manage the configuration | | | X | X | |
| | X | Manage problems and incidents | | | X | X | X |
| | X | Manage data | | | X | X | |
| X | | Manage facilities | | X | | | |
| | X | Manage operations | | | X | X | |
| | | **Monitoring** | | | | | |
| X | | Monitor the process | | | | X | X |
| X | | Assess internal control adequacy | | | | | X |
| X | | Independent assurance | X | | | | X |
| X | | Provide for independent audit | | | | | X |

Source: IT Governance Institute (2004), Page 50.

**Figure 6: COSO, CobiT, 17799, Common Criteria, and ITIL**

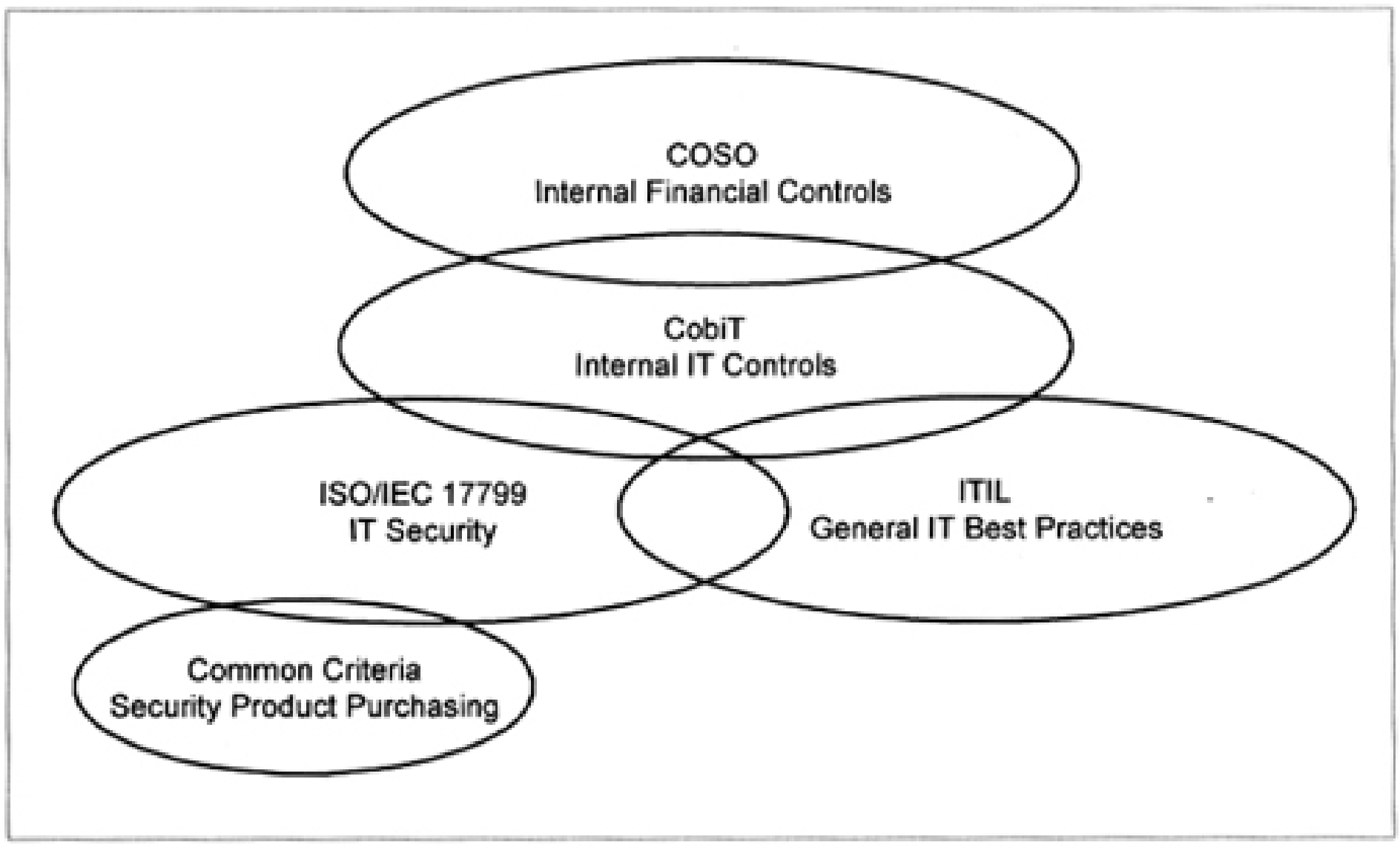

**Figure 7: Testing Methods for Material Financial Spreadsheets**

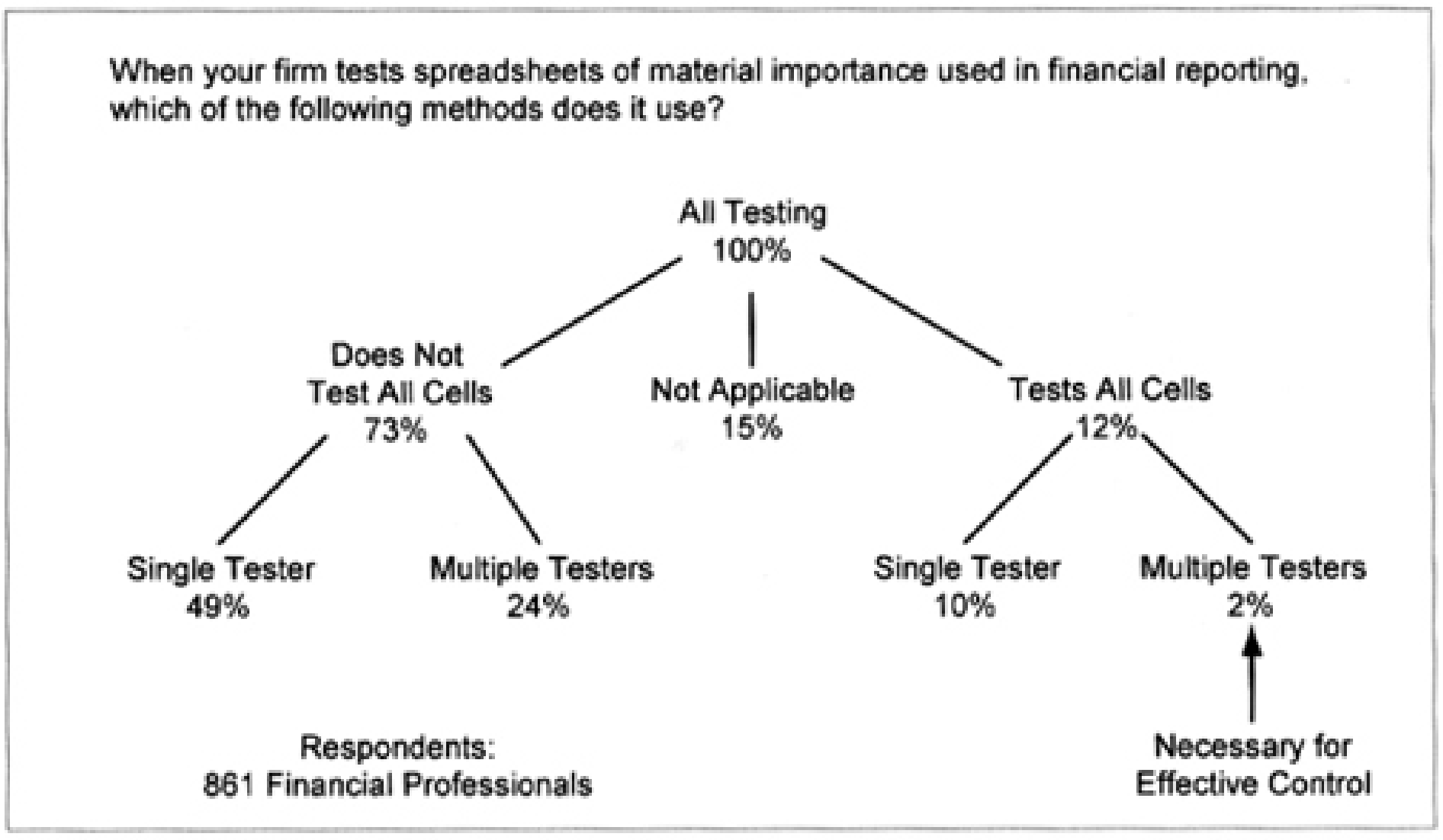

**Figure 8: The Fraud Triangle**

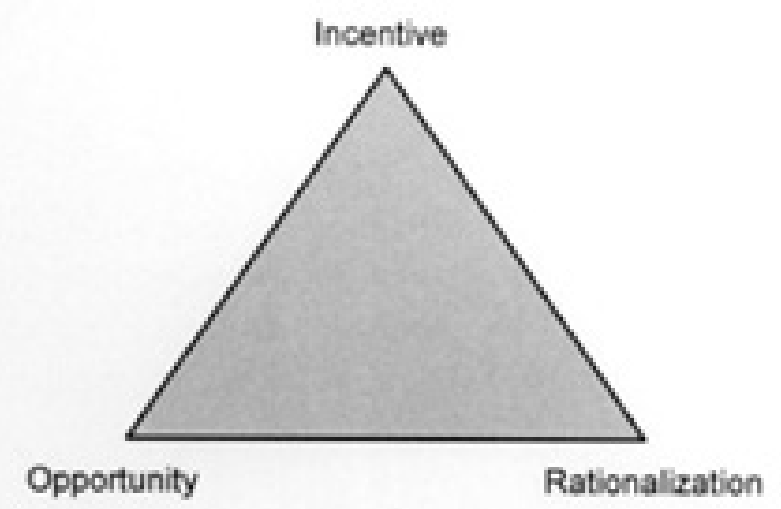

Source: Accounting Standards Board (2002) SAS 99

# Figures